\documentclass[11pt,A4,fleqn]{article}
\usepackage{graphicx}
\usepackage{amssymb}
\usepackage{amsmath}
\usepackage{subfigure}
\usepackage[top=0.5in]{geometry}

\begin{document}

\title{Raft Consensus Algorithm: an Effective Substitute for Paxos in High Throughput P2P-based Systems 
}


\author{Mahmood Fazlali         \and
        Sara Moazezi-Eftekhar    \and
        Mohammad Mahdi Dehshibi  \and
        Hadi Tabatabaee Malazi   \and
        Masoud Nosrati
}




\maketitle

\begin{abstract}
One of the significant problem in peer-to-peer databases is collision problem. These databases do not rely on a central leader that is a reason to increase scalability and fault tolerance. Utilizing these systems in high throughput computing cause more flexibility in computing system and meanwhile solve the problems in most of the computing systems which are depend on a central nodes. There are limited researches in this scope and they seem are not suitable for using in a large scale. In this paper, we used Cassandra which is a distributed database based on peer-to-peer network as a high throughput computing system. Cassandra uses Paxos to elect central leader by default that causes collision problem. Among existent consensus algorithms Raft separates the key elements of consensus, such as leader election, so enforces a stronger degree of coherency to reduce the number of states that must be considered, such as collision.
\end{abstract}

\section{Introduction}
According to the CAP theorem, consistency, high availability, and partition tolerance are the three axes of any distributed database management system that at most two of them can be fulfilled. The systems with the primary goal of scalability usually provide high availability and partition tolerance. Therefore, they have to use various approaches to managing the replicas, and also be prepared for different partitioning scenarios. Horizontal partitioning in server clusters is one of the cost-effective ways for providing scalability. By horizontal partitioning, each cluster manages a part of data with its own database system. In such a system, manual processes are dedicated to maintenance and load balancing tasks; but recently, new architectures emerged for automatic horizontal partitioning and load balancing~\cite{ref2,ref3,ref4,ref5} that are mostly based on hash keys. On the other hand, accessibility is the second challenging issue, which is almost achieved by replication. 

The simplest solution for replication is a master-slave relationship in pairs of nodes. As a matter of fact, many database systems implement the horizontal partitioning by master-slave replication. But, this solution is accompanied with some limitations. For instance, if one of the nodes fails to continue the execution, the other one can still do the tasks, but in the case of existing, a sequence of failures this system will be in trouble. Another solution is tripartite replication that is more reliable and facilitates the management of problematic situations. Also, it provides the offline update of replica, while two other ones are online~\cite{ref2}. Managing the sequence of failures is more convoluted in this solution. So, many efforts aimed at simplifying and solving this problem~\cite{ref6}. Experiments disclosed that the best way for managing the replicas that are more than three in numbers is utilizing consensus algorithms~\cite{ref7}. Peer-to-peer (P2P) database systems are the most prominent example of this type of systems.

A P2P database is based on a network of heterogeneous interconnected nodes for sharing data via the files, in the role of both supplier and consumer. In this structure, the function of all the nodes is the same, and all the nodes are autonomous. Theoretically, there is no central coordinator in the system, but in practice, it is not materialized yet. In fact, peers construct an overlapping network and form a topology for connections that is able to put the nodes in touch after joining and keep the links for rejoins~\cite{ref1}. 

Cassandra is one of the outstanding products that is based on the hash table and has a P2P structure~\cite{ref3}. It is an open source distributed database management system that is developed based on Amazon Dynamo and Big table data model. Some of the key features are distribution, decentralization, scalability, high accessibility, fault tolerance, tunable stability and column-oriented model. Cassandra is an appropriate tool for managing the vast amount of data on the structure of an intricate network. Some other features such as compatible hash/fragmentation, a rumor-based algorithm for controlling the membership, replication, anarchy algorithm, and failure detection made Cassandra be a distributed storing system, which is fault tolerant and symmetric total compatibility. 

Ultimate compatibility that is provided by Cassandra causes some unexpected problems. For example, when different data are written in Cassandra, multiple copies in different timestamps are created. The major issue emerges when some of the nodes choose an identical item from the queue for running. So, one of them takes the job and omits it from the queue, while for the other nodes it might be thought that the job is still in the queue, and they try to run it. This situation is called workers’ collision~\cite{ref2}. Consensus algorithms, such as Paxos~\cite{ref54}, are proposed to solve this problem. They let the applicant nodes to be noticed about preceding candidates that have the priority for popping the jobs~\cite{ref2}.

Paxos is a group of protocols for performing consensus in a network of unreliable nodes. The main limitation of Paxos is complexity in processing and difficulty in understanding the details that entail inappropriate architecture of a system in practice. Proposed solutions were directed to emerge Raft algorithm~\cite{ref56}, which was simpler than Paxos~\cite{ref4}. Raft utilized concepts like leader election, log, and security to make the consensus more comprehensible. Raft is equivalent to the Paxos in fault-tolerance and performance, but it separates the consensus into relatively independent sub-problems, and clearly addresses all essential pieces for real-world implementation of the system.

Raft starts the consensus with selecting a leader and giving the authority of managing the logs to it. Raft provides the fault-tolerance when the leader takes the logs from the servers and gives it to others to let them know about the secure time of receiving logs. Previous experiments indicated that Paxos has more latency in reading and writing.

In this paper, we will briefly show the similarities and differences between Paxos and Raft. Firstly, we will describe what consensus algorithms problem is. Secondly, we will describe other Consensus algorithms and protocols. We will describe how leaders are elected in  Paxos and Raft algorithms Finally, the goal of our experiment is to find a solution for collision problem in distributed queue for Cassandra database system; because it has a positive correlation with the efficiency of high throughput capabilities and reducing the redundancy of P2P storage system. Due to it, some of the well-known consensus algorithms are surveyed involving: Paxos, Viewstamped Replication (VR), Zab, Chandra-Toueg, and Raft. 

Studying these algorithms entails focusing on Paxos and proposing the usage of Raft solution in Cassandra, to manage the replicas in a more optimized manner and decrease the read/write request latency for achieving better performance and fairer load balancing.

In the rest of this paper, we first define the consensus problem, then in Section \ref{sec.cons-alg} an inspection of previous work is presented. This claim is underpinned in Section \ref{sec.exp} by providing the results of simulation and different assessments. Finally, in Section \ref{sec.con} we get into the conclusions.

\section{Consensus Problem}
\label{sec.cons}
In this section, we take a look at the basic concepts of consensus problem.

Consensus deals with the consenting on a similar data by some processes; whilst some of these processes might be failed or unreliable. Hence, consensus algorithms must be able to handle the failures. The general approach is that a common data consents. A faulty process might change the results of consensus to a wrong value. Consensus algorithms must have the following features to tolerate the temporary failures:

\begin{itemize}
\item	Termination: Each valid process terminates on finite values.
\item	Validity: If all the processes propose the real value of $v$, all the valid processes choose $v$.
\item	Integration: Each valid process decides on at most one value.
\item	Agreement: Each valid process must consent on a common value.
\end{itemize}

Besides, the system might encounter with two types of failures:

1) Crash failure: It occurs when a process stops abnormally.

2) Byzantine failure: A process with Byzantine failure might send wrong data to other processors, or hibernate for a long time.

The second type of failure is more destructive and harder to handle. So, it is expected that the consensus algorithms anticipate the occurrence of such failures and attempt to neutralize their effect.

\section{Consensus Algorithms and Protocols}
\label{sec.cons-alg}
A brief review of the prominent consensus approaches is provided in this section to investigate the one which best fits real-world application on Cassandra database system. But first, some of the required backgrounds is presented.

A distributed system is comprised of a collection of independently interconnected computing devices that seems like a single coherent system to end user~\cite{ref8,ref9}. This concept is generalized for distributed database systems, too. Distributed database systems are applications that provide the accessibility to data in a distributed environment. Designers of such these systems attempt to provide the fault tolerance by techniques like replication and data distribution; but, these techniques are accompanied by increasing the redundancy and difficulty in management and implementation~\cite{ref10}.

As it was mentioned, the intended database system in this research is Cassandra, which is designed by Avinash Lakshman and Prashant Malik in 2008, and it was initially utilized by Facebook. Cassandra is an open source database system based on the Amazon Dynamo, for managing massive data in the Big table model.

In the following, we will review the literature and survey some of the well-known consensus algorithms.

\subsection{Background}
According to~\cite{ref11,ref12}, P2P systems were devised for sharing the files. Calculations in P2P systems led to exuding excellent research areas, like issues related to searching~\cite{ref13,ref14,ref15}. It attracted the attention of many studies to this scope. In~\cite{ref16}, it was stated that a P2P system has the following features: 

\begin{itemize}
\item It is absolutely distributed, without central leading points.
\item Users are able to enter the tasks from their connected node and then disconnect from the network.
\item Global scheduling is needed.
\end{itemize}

Some of the extant studies such as~\cite{ref17,ref18,ref62} utilize the centralized solutions and cannot provide the complete distribution. Other works such as~\cite{ref19} and~\cite{ref20} use the super-peer model that is rather distributed and lets the worker nodes to be connected and scheduled by super-peers. In~\cite{ref21} a distributed scheduling is proposed based on the overlay network and tree structure, but in that system, the tasks must be supervised. Hence, the second requisite is violated. Centralized leadership as the most important drawback of~\cite{ref22} and~\cite{ref23} imposes limitations in scalability and accessibility to high throughput systems. For coping with it, consensus algorithms are proposed that let the processors transfer their local calculation results to each other to gain coordination~\cite{ref24,ref25}. Establishing simple local rules can cause the consonant behavior of nodes~\cite{ref26,ref27} to a same direction and convergence of their acts~\cite{ref28}. In one of the prominent research in this field, a special class of consensus algorithms was proposed based on the convergence of weak connections~\cite{ref29}, and it initiated following studies like~\cite{ref30} for weakly connected graphs and~\cite{ref27} for asynchronous systems~\cite{ref31}.

Different replicas in the system might see the multiple copies of events with different sequences. This problem is addressed as \emph{casualty} in many studies like~\cite{ref32} by Lamport. In~\cite{ref33,ref34,ref35} some solutions were presented, but no one was perfect. Lamport in~\cite{ref36} claimed that it is possible to achieve to consent even with some faulty processes. But, asynchronous systems are out of this rule~\cite{ref37}.

\subsection{Paxos protocol}
Paxos is known as the most popular consensus protocol, which was presented by Lamport’s paper~\cite{ref37} and developed by Lampson’s article~\cite{ref38}. Paxos is a consensus protocol for a network of unreliable processors. The consensus is the process of consenting on a particular value or item among a group of participants. This process is hardened when there are failures in processing or communication. Besides, Paxos guarantees the compatibility by applying a broad range of assessments on processors, latency, level of participation of nodes, messages, and failures. On the other hand, Paxos has an intricate structure that makes it inappropriate for practical system design.

There exist many real-world implementations of Paxos such as Chubby~\cite{ref39}, Megastore lock service~\cite{ref40} and Spanner storage system~\cite{ref41} that are created by Google, Autopilot service~\cite{ref42}, Bing, Azure~\cite{ref43} by Microsoft, Ceph storage system.

Two significant optimizations for modifying the performance of Paxos are batching~\cite{ref44} and pipelining~\cite{ref45}. Batching is used for message aggregation based on TCP's Nagle algorithm, and pipelining is a general technique for optimization that runs the requests in parallel to improve the efficiency. An example of pipelining is HTTP~\cite{ref46}. With a brief look at literature, it is inferred that no effort is made to blend batching and pipelining techniques, so far. But, some works were conducted to optimize the Paxos based protocols. For example, LCR in~\cite{ref47} proposes a pervasive atomic protocol based on the ring topology for optimizing clock vectors. Ring Paxos~\cite{ref48} combines different techniques for improving the efficiency of networks such as IP multicast, ring topology and a minimum number of acceptors. Both~\cite{ref47} and~\cite{ref48} just targeted the LAN environment; hence they utilize LAN specific techniques like IP multicast.

Although Paxos is widely used for consensus in recent studies, it has some shortcomings that are forked from its single-decree formulation structure. That is, the servers can only consent on one of the entries in the log. It causes hardship in understanding the consensus mechanism, because it consists of two phases that cannot be separated, and also Paxos uses the weak leader for optimization which is not symmetric, yet.

\subsection{Paxos algorithm}
Paxos lets a group of processes (called replicas) to consent on a similar value, in a faulty environment, which replicas might fail, and messages might lose in the network. Each replica is able to retrieve the previous state before the failure. If all the replicas perform without failure for a particular time period, Paxos algorithm can guarantee the similarity of consented value among the processes. Paxos contains three steps that would be repeated in the case of existing failures~\cite{ref53}:

1) Choosing a replica as the leader

2) The leader chooses a value and sends it to all other replicas in the form of an \textit{accept} message. Other replicas can accept (by \textit{ack} message) or reject (by \textit{reject} message) the value.

3) If the majority of replicas consent with the leader, the consensus is done, and the leader sends the \textit{commit} message to all the replicas.

In Paxos, several processes might decide to be the next leader. So the system dominant policies should restrict them to compete because it makes delay in carrying on the consensus. The flexibility that is caused by multileader imagination leads to selecting multiple values, but it cannot detect the results. In fact, Paxos guarantees the single value consensus via the attaching order number to the leaders and restricting the leaders’ options for selecting the values. Putting the leaders into the order lets the replicas distinguish the current leader from previous ones that entails ignoring the messages from former leaders and accepting the messages of the current leader.

\begin{figure}
\centering
\includegraphics[width=0.75\columnwidth]{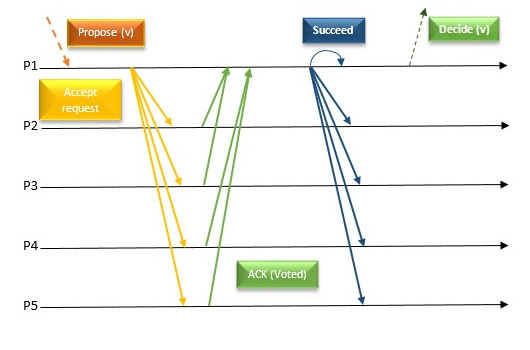}
\caption{Paxos algorithm~\cite{ref54}.}
\label{fig1}
\end{figure}

A replica sends a \textit{proposed} message to others  whenever it decides to be the new leader. If other replicas accept the proposed leader, they \textit{promise} to reject the messages from previous leaders~\cite{ref53}. More details about it can be found in~\cite{ref54}. Figure \ref{fig1} shows an example of it.

Results of comparing the Chandra-Toueg and Paxos that are presented in~\cite{ref53} show that Paxos has a better performance in the case of existing failed leader, but in other situations, the latency -and consequently the performance- are almost the same for both of algorithms.

\subsection{Chandra-Toueg}
Chandra-Toueg~\cite{ref53} is presented for a network of unreliable processes and utilizes strong failure detector. Failure detector is a brief version of timeouts. This algorithm sends a signal to each process to detect the failures; so it could be regarded somehow similar to Paxos. Both of them are based on the strong failure detector, and the number of their failed processes is less than n/2, where n is the total number of processes.

\begin{figure}
\centering
\includegraphics[width=0.75\columnwidth]{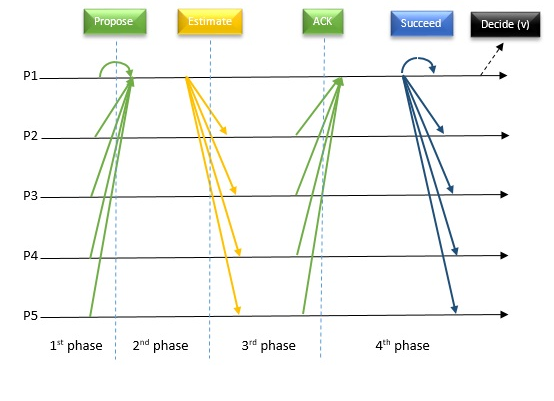}
\caption{Phases in a round of Chandra-Toueg~\cite{ref53}.}
\label{fig2}
\end{figure}

The algorithm proceeds in rounds which are shown in Figure \ref{fig2} and utilizes a rotating coordinator: in each round r, the process whose identity is given by (r mod n)+1 is selected as the coordinator. Each process keeps track of its current preferred decision value (initially equal to the input of the process) and the last round that it altered its decision value (the value's timestamp). The actions in each round are:

1) All processes send ($r$, $preference$, $timestamp$) to the coordinator.

2) The coordinator waits to receive messages from at least half of the processes (including itself). It then chooses as its preference a value with the most recent timestamp among those sent.

3) The coordinator sends ($r$, $preference$) to all processes.

4) Each process waits to receive ($r$, $preference$) from the coordinator or for its failure detector to identify the coordinator as crashed. In the first case, it sets its own preference to the coordinator's preference and responds with $ack(r)$. In the second case, it sends $nack(r)$ to the coordinator.

5) The coordinator waits to receive $ack(r)$ or $nack(r)$ from a majority of processes. If it receives $ack(r)$ from a majority, it sends $decide(preference)$ to all processes.

6) Any process that receives $decide(preference)$ for the first time sends $decide(preference)$ to all processes, then decides preference and terminates.

\subsection{Zab}
Zab is a crash-recovery atomic broadcast algorithm designed for the ZooKeeper coordination service. ZooKeeper implements a primary-backup scheme in which a primary process executes clients operations  and uses Zab to propagate the corresponding incremental state changes to backup processes~\cite{ref60}.

Zab at a high level is a leader based protocol similar to Paxos. Some of the aspects that distinguish Zab from Paxos is that Zab recovers histories rather than single instances of protocols; Zab has prefix ordering properties to allow primary/backup implementations to have multiple outstanding operations outstanding at a time; and Zab does special processing at leadership change to guarantee unique sequence numbers for values proposed by a leader. All servers start off looking for a leader. Once the instance of leader election at a given server indicates a leader has emerged it will move to phase 1. If the leader election instance indicates that the server is the leader, it will return to Phase 1 as a leader, or do as a follower.

\textbf{Phase 1}: In this phase, an elected leader makes sure that previous leaders cannot commit new proposals and decides on an initial history. (Note a leader is also considered a follower of itself.)

\textbf{Phase 2}: Sync with followers

\textbf{Phase 3}: The leader and followers can have multiple proposals in the process and at various stages in the pipeline. The leader will remain active as long as there is a quorum of followers acknowledging its proposals or pings within a timeout interval. The follower will continue to support a leader as long as it receives proposals or pings within a timeout interval. Any failures or timeouts will result in the server going back to leader election. 

\subsection{Viewstamped Replication}
VR provides state machine replication in an asynchronous network like the Internet and handles nodes failures. VR code contains logic for maintaining a consensus among the majority of replicas for each operation. VR proxy sits between the real application clients and handles its request. It shares the configuration state such as current view number, current primary with replicas. The VR protocol can be broadly divided into three modes of operations:

1) \textbf{Normal Operation}: In this mode, the primary replica accepts an operation from the VR proxy and queues them for replicating onto at least f+1 replicas (including itself). Once the operation is logged on majority of replica, it is committed on the primary and is executed against the application service and the response is updated to client. 

2) \textbf{View Change}: If the primary goes down, the replicas will have timeout alerts and will start the view change algorithm. Each replica will send StartViewChange to all replicas, and when a replicas f get such requests, it sends a DoViewChange message to the new primary.
The new primary waits for f+1 DoViewChange message and picks the most updated logs from this call. It then sends StartView to all the replica and resume operation in normal mode again. Since log state is transferred in multiple messages, this phase has high throughput. Checkpointing the logs will be used to decrease the amount of exchanged bytes.

3) \textbf{StateRecovery}:When a replica is joined back after a failure, it catches up with the latest log through.

\subsection{Raft}
The Raft is designed to facilitate the understanding of consensus problem by dividing it to different sub –problems to get to better quality in consensus~\cite{ref55}. It would be helpful to take a look at the concept of distributed consensus, to have a clear perception of Raft.

In a system with multiple processes, each process might be at one of these states: follower state, candidate state or leader state. All the processes are initially at the follower state. If they do not receive a message from the leader, they can go to candidate state. Candid process requests for votes from other processes and the candid with the majority of votes becomes a leader. This process is called leader election. Then, all the changes of the system are supervised by the new leader. For the consensus, leader waits for the majority of processes to write their entries in the log and after the completion of the log, leader notifies the followers about the completion and consensus result~\cite{ref55}. 

Given the leader approach, Raft disjoints the consensus problem into three independent sub-problems as follows:

\begin{itemize}
\item Leader election: a new leader must be chosen when an existing leader fails.
\item Log replication: the leader must accept log entries from clients and replicate them across the cluster, forcing the other logs to agree with its own.
\item Safety: the key safety property for Raft is the State Machine Safety Property: if any server has applied a particular log entry to its state machine, then no other server may use a different command for the same log index.
\end{itemize}

As the last notion, it must be stated that Raft is similar to Paxos from a structural point of view. But, it is simpler and more understandable, and it can cope with the problem of read/write request latency. In the next section, we will show that Paxos can be replaced by Raft in Cassandra database system to achieve better performance. Experimental results can prove our claim. 

\subsection{Discussion}
Among the existing consensus algorithms, Raft can be compared with Paxos, VR, and Zab; partially because all of them can handle the fail-stops. VR and Zab are leader based algorithms that are more similar to Raft. Both of them choose a leader at initial step, and the leader handles the replicated logs. Their difference is in the managing the leader and maintenance operation of logs. On the other hand, Raft has a lighter mechanism rather than VR and Zab. For example, VR and Zab use 10 types of messages, while Raft just utilizes 4 types (two RPC requests and their answers). But, Zab provides a stronger guarantee for concurrent running of requests by pipelining, and also observes the FIFO order of them~\cite{ref49}. Furthermore, there is no failure detector in Paxos, VR, and Zab. It could be inferred that failure detection process is considered absolutely disparate from consensus, while Raft considers the failure detection via the use of heartbeats. Another notion is about scheduling. The aforementioned algorithms allocate the number of turns to the servers in two ways: Zab and Raft do the consensus to be sure that there is a unique leader in each time. If system nodes consent on a server, then it can replicate the log entries. But, Paxos and VR divide the turns’ number domain so that the servers cannot compete for turns (like Round-Robin fashion). In practice, there is no difference in the application of these methods, because both of them do the consensus, accurately.

Table 1 shows some details of the different consensus algorithms. The field \emph{New leader} shows that which servers are able to be selected as next leader; \emph{Management of configurations} indicates the ability of algorithm in configuring the server for leadership during the selection process; and finally \emph{Vote collector} shows the servers that can perform the consensus  and collect the votes~\cite{ref11}.

\begin{table}[!t]
\centering
\footnotesize\setlength{\tabcolsep}{4pt}
\caption{Details of leadership in different algorithms}
\label{tab:tbl1}
\begin{center}
\begin{tabular}{|l |c |c |c|}
\hline
Algorithm & New leader & Management  & Vote collectors\\
               &		&of configurations&\\
\hline
Paxos~\cite{ref37} &   & & \\
\hline
Paxos~\cite{ref32} & All the servers   & Yes & New leader\\
\hline
Chandra-Toung~\cite{ref53} &   &  &\\
\hline
VR~\cite{ref51} & Servers with& No & View manager\\
  & most recent logs &  & \\
\hline
VRR\textsuperscript{*}\cite{ref59}  & Determined by  & No & New leader\\
  & view number  &  & \\
\hline
Zab\cite{ref49} & Servers with & No  & New leader\\
&most recent logs  &  & \\
\hline
Raft~\cite{ref56} & Servers with & No  & New leader\\
 &most recent logs  &   &\\
\hline
\end{tabular}
\end{center}
* Viewstamped Replication Revisited utilizes a round turn approach for leader selection.
\end{table}

Paxos (non-optimized version), Zab and VRR (VR) need more process to ensure that the new leader has the complete list of entries; because they disregard the logs in choosing the new leader. In Paxos, a leader runs two single-decree phases for each log entry, and it is not aware of the completion of the log. It might cause a long latency period until choosing a new leader. Zab and VRR submit the whole logs to a new leader, and new leader wants the most recent one. This is a good idea, but difficult in practice. It is recommended to decrease the entries to gain more optimization.

As the last notion, there are various ways to reduce the number of servers, while it does not detect the fault tolerance. A common way is simplifying the replication so that it uses few number of servers in a cluster~\cite{ref63}. This approach is called “thrifty”~\cite{ref50}. It decreases the network load to half of normal situation because it replicates the entries for half of the servers in the cluster. Other methods are proposed in~\cite{ref51} and~\cite{ref52} using witness servers, that are out of the scope of current paper to be surveyed. Due to the aim of this article that is based on analysis of fault-tolerance detection which affect the results. 

One of the most significant factor in consensus algorithms performance is leader election which is compared in Paxos and Raft in the following paragraphs.

Both Paxos and Raft assume that eventually there will be a leader that all stable servers trust and a single leader is responsible for a term. A new leader will propose a new term, which must be greater than the previous one, if the current leader is suspected to have failed.

In Paxos leader election procedure is unconditionally, the only considerable thing is that a server which is a candidate must be stable and available. These conditions are also the same about Raft, but there is one effective difference that a server which is a candidate as a leader, must be the most up to date server among others. In Paxos because any server deserves to become a leader, since a cluster which contained the leader received a request data for the past that the leader does not have any background about that, must learn it from the other servers. This causes more pressure in the cluster. Whereas, this problem is solved in Raft with its enhanced architecture by choosing the most up to date server as a leader.

It would be noticeable that the leader in both Raft and Paxos can affect the traffic goes through it.

\section{Experimental results}
\label{sec.exp}

To compare the Paxos and Raft, four evaluation parameters are considered as follows:

1) I/O latency: It refers to the speed of read/write operations. Latency is the time duration between the start point and end point of a transaction. For database systems, it is so important to achieve less latency because of a vast number of I/O operations that are needed for executing the transactions.

2) Read/Write requests: Load balancing must be considered as an important factor when clustering the nodes to avoid of latency. It should be stated that this issue is observed in Cassandra via replication and providing similar data for nodes. Based on the default settings, several queues are utilized for read/write operations. It causes load balancing on one side and evading of collisions on the other side. Also, it lets the system accept the clients' requests even during the maintenance time.

3) OS load: In Cassandra, each node is potent enough to utilize the 100\% of processor's power. OS load refers to the imposed load to the hardware through the operating system.

4) OS: Net sent/received: It relates to the amount of data that is transferred via the network by the operating system. Network load is selected as a measure for assessment of consensus algorithms, because of the huge amount of read/write operation load that Cassandra keeps in the queues.

For experiments, we utilized a Cassandra server with following details:

- CPU: Intel Xeon X5660 @ 2.80 GHz with 6 Cores and 12 threads

- Memory: 32 GB, DDR3

We installed four virtual machines with Linux (Ubuntu 15.04) and Cassandra database in the aforementioned system. To monitor the virtual machines we utilized Opscenter (from DataStax.com). Also, Devcenter (from DataStax.com) was used for establishing the connection with the database and managing the queries. At the next step, a cluster was created that was comprised of all four virtual machines, which we may call nodes. The nodes were organized in a ring topology. To evaluate the system in high network load situation, stress tool was utilized. This is an application that auto generate a vast amount of data for Cassandra and other databases to analyze their behavior. In this case, is used Cassandra stress tool from DataStax.com.
\begin{figure}[!t]
\begin{small}
\begin{center}\footnotesize
\subfigure[]{ \includegraphics[scale=0.49]{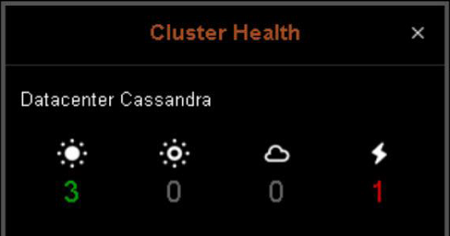}}
\subfigure[]{ \includegraphics[scale=0.49]{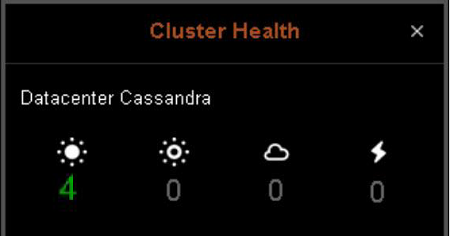}}
\caption{Cluster health under the high load using Paxos (a) and Raft (b) consensus algorithms.}
 \label{fig4}
\end{center}
\end{small}
\end{figure}
The specifications of the nodes:

- Number of Nodes: 8 in WMW

- Hard Disk: 10 G

- CPU: 2.8 GHZ Intel

- Memory: 2G

- OS: Ubuntu 15.04 64X 

The test parameter: ~/Cassandra-stress/target/appassembler\# bin/stress –o insert –b 1000 –c 10 –n 9000000 –t 5 192.168.142.172:9160.

During the experience with Paxos, we increased the load continuously. In such a circumstance, one of the nodes stopped, while the others were interacting under a high load. Stopped node was not able to join again and stayed out.

In the experience of using Raft, we installed golang-raft-dev package. Also, Opscenter and Stress tool were utilized as well as the experience of Paxos.

Unlike Paxos, in Raft experience it was observed that the cluster health was improved, and none of the nodes stopped by increasing the load. Also, all of them had a normal load situation. This point can be considered as the superiority of Raft rather than Paxos.
\begin{figure}[!t]
\begin{small}
\begin{center}\footnotesize
\subfigure[]{ \includegraphics[scale=0.40]{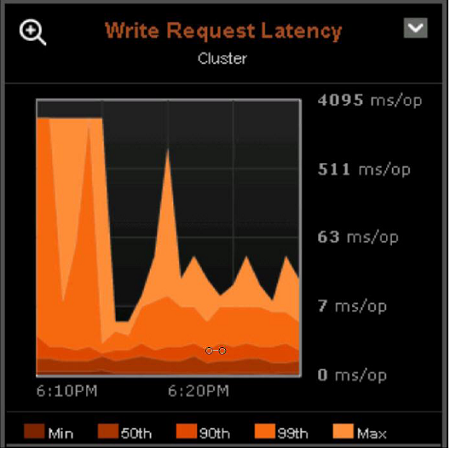}}
\subfigure[]{ \includegraphics[scale=0.40]{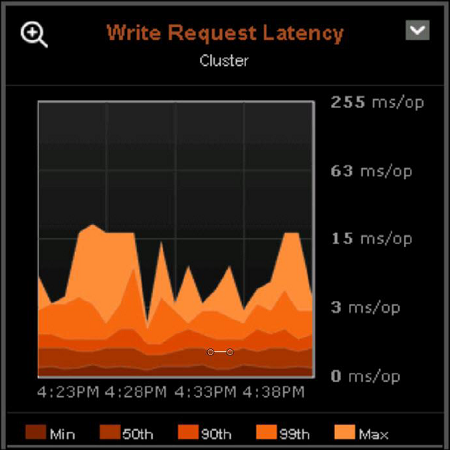}}
\caption{Write request latency in Paxos (a) and Raft (b).}
 \label{fig5}
\end{center}
\end{small}
\end{figure}

\begin{figure}[htb]
\begin{small}
\begin{center}\footnotesize
\subfigure[]{ \includegraphics[scale=0.45]{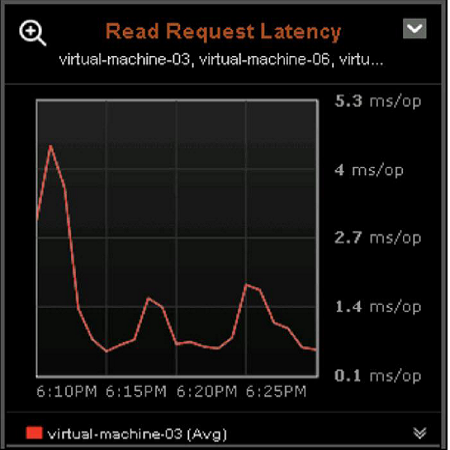}}
\subfigure[]{ \includegraphics[scale=0.45]{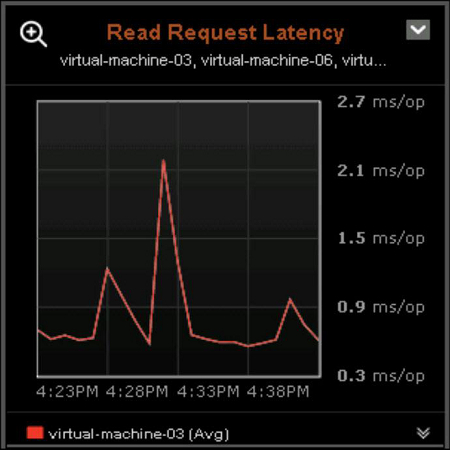}}
\caption{Read request latency in Paxos (a) and Raft (b).}
 \label{fig6}
\end{center}
\end{small}
\end{figure}
Figure \ref{fig4} shows the cluster health in Paxos and Raft. In Paxos experiment, by increasing the load, one of the nodes encounter with the problem. It is evident that three servers are operating under a massive load of the system, while one of the nodes is down and it cannot return to the process again. But, Raft shows more flexibility by increasing the load. All four servers are operating under more balanced load, accurately. 

It can be seen from the figures that Raft can divide load pressure among the nodes in the cluster. So that, the nodes are more relaxed in compare with using Paxos.

As it is mentioned before, Raft performance to conserve mechanisms of cluster by choosing proper leader is far better than Paxos in omitting pressure on the other servers. As it can be seen from the Figure \ref{fig11}, this case can be easily sensed.

The vertical axis shows the write requests operator performance in a millisecond, and the horizontal axis demonstrates the period of time. Write request latency is provided in Figure \ref{fig5} that denotes on unbalanced write request latency of system in Paxos, and better performance of Raft. By increasing the load, write request latency in Paxos reaches to about 4000 ms/op, while the maximum latency in Raft is about 20 ms/op. Also, the minimum of Raft is approximately half of the Paxos. These differences denote on acceptable optimization of write request latency in Raft rather than Paxos. These figures illustrate the writing operation in Raft is better than Paxos because it is vividly obvious that in Paxos write requests are likely buffered and wait to write, therefore aggregation of write requests will be increased. However, this situation in Raft is so fewer.

\begin{figure}[htb]
\begin{small}
\begin{center}\footnotesize
\subfigure[]{ \includegraphics[scale=0.45]{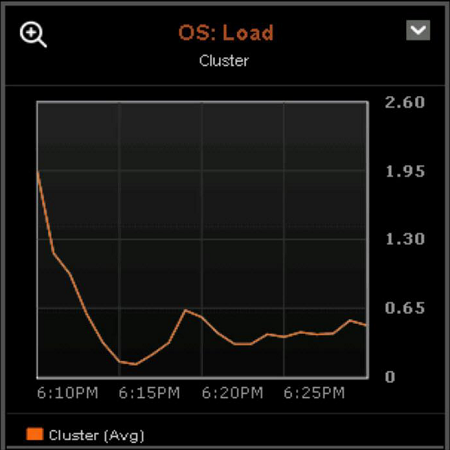}}
\subfigure[]{ \includegraphics[scale=0.45]{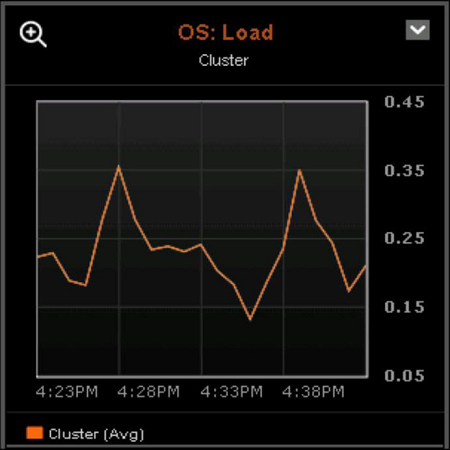}}
\caption{Operating system load in Paxos (a) and Raft (b).}
 \label{fig7}
\end{center}
\end{small}
\end{figure}

\begin{figure}[htb]
\begin{small}
\begin{center}\footnotesize
\subfigure[]{ \includegraphics[scale=0.45]{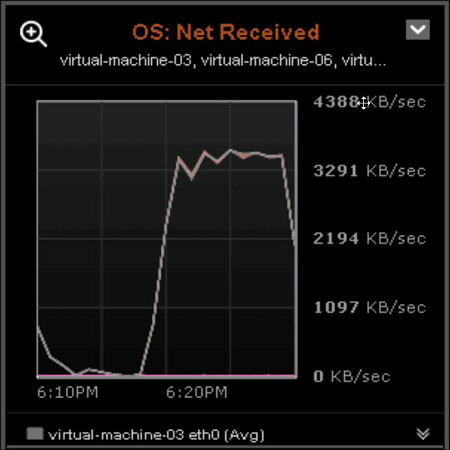}}
\subfigure[]{ \includegraphics[scale=0.45]{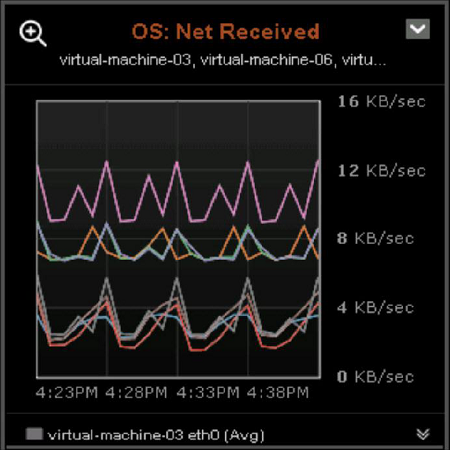}}
\caption{Network received data by the operating system in Paxos (a) and Raft (b).}
 \label{fig8}
\end{center}
\end{small}
\end{figure}

\begin{figure}[!t]
\begin{small}
\begin{center}\footnotesize
\subfigure[]{ \includegraphics[scale=0.45]{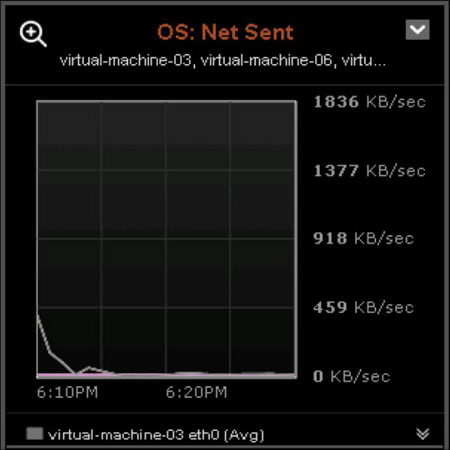}}
\subfigure[]{ \includegraphics[scale=0.45]{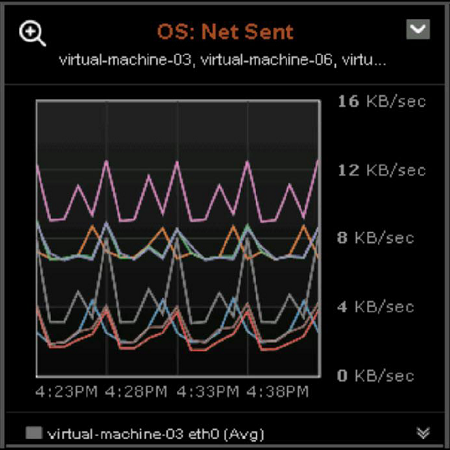}}
\caption{Operating system load in Paxos (a) and Raft (b).}
 \label{fig9}
\end{center}
\end{small}
\end{figure}

\begin{figure}[htb]
\begin{small}
\begin{center}\footnotesize
\subfigure[]{ \includegraphics[scale=0.45]{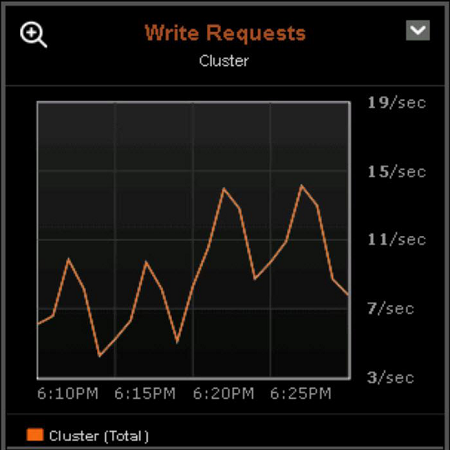}}
\subfigure[]{ \includegraphics[scale=0.45]{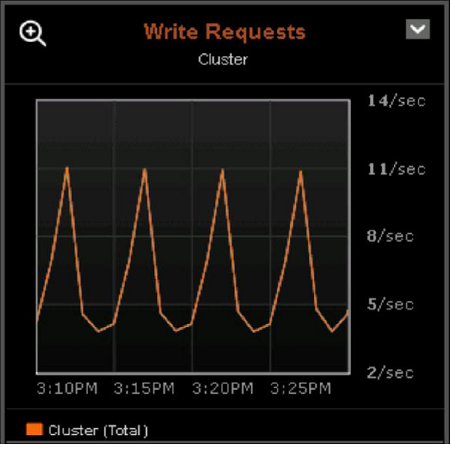}}
\caption{Write requests in Paxos (a) and Raft (b).}
 \label{fig10}
\end{center}
\end{small}
\end{figure}

\begin{figure}[htb]
\begin{small}
\begin{center}\footnotesize
\subfigure[]{ \includegraphics[scale=0.45]{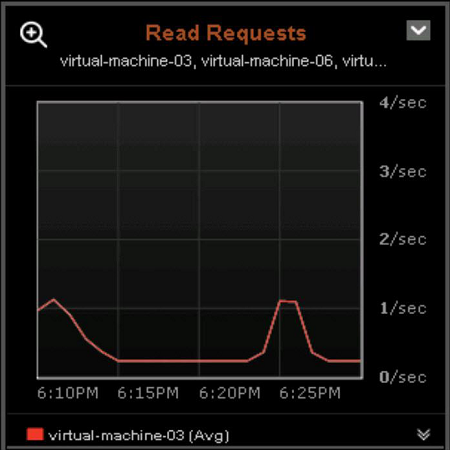}}
\subfigure[]{ \includegraphics[scale=0.45]{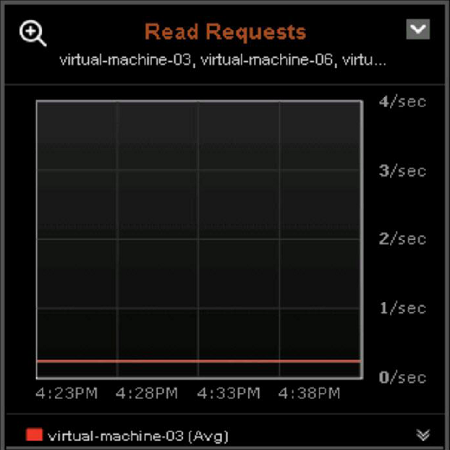}}
\caption{Read requests in Paxos (a) and Raft (b).}
 \label{fig11}
\end{center}
\end{small}
\end{figure}
The vertical axis shows the write requests operator performance in a millisecond, and the horizontal axis shows the period of time. Figure \ref{fig6} shows the read request latency for one of the nodes in the cluster as a sample. It is clear that in Paxos, by increasing the write request latency, read request latency soars. It could be inferred that read and write request latency are interdependent. On the other hand, read request latency in Raft still follows a more balanced pattern. Also, the maximum value for latency in Paxos that is about 4.5 ms/op is rather greater than Raft, which is approximately 2.2 ms/op. Although read request latency is much smaller than write request latency, this read request latency can have an impressive effect on the overall speed of the system, especially in the case of larger clusters. These figures illustrate the read requests latency in Paxos is twice more than Raft.

Figure \ref{fig7} indicates the average operating system load of the cluster. Maximum loads of Paxos and Raft are about 2 and 0.4, respectively, that shows the efficiency of Raft in comparison with Paxos. Besides, comparing the operating system load with read/write request latency indicates that both of them recess by increasing the system load. So, it could be inferred that there is a positive relationship between OS load and read/write request latency. But, in Raft there is not such a correlation. On the other hand, OS load in Paxos changes in a wider domain, while Raft follows a more regular pattern in its changes. It shows the appropriate load balancing of Raft rather than Paxos. These figures demonstrate that in the case of using Raft average of OS load in the cluster not only is incredibly less than Paxos, but it is also more monotonous.

In Figure \ref{fig8}, the vertical axis shows the rate of operating data received in the cluster base on KB/sec while the horizontal axis shows the period of time. 

In Figure \ref{fig9}, the vertical axis shows the rate of operating data sent base on KB/sec in the cluster while the horizontal axis illustrates the period of time. 

Figures \ref{fig8} and \ref{fig9} show the network sent and received load for one of the nodes in the cluster. In Paxos, received data has an inverse relationship with read/write request latency. The reason could be described easily as by increasing the latency, nodes send less data to the network. Consequently, both sends and received slake when the load is high. Posted data converges to zero when read/write request latency is reaching to the maximum value. But, sent and received data in Raft have steadier rate and more regular pattern that results from appropriate load balancing in this system. It is obvious from the figures 8 that the maximum network received in Raft in compare with Paxos is surprisingly far less and more monotonous and also as is evident from the figures 9 that the maximum transfer rate is in the cluster of  Raft in compare with Paxos is more monotonous.

The vertical axis shows the number of write requests in a second in the cluster and the horizontal axis demonstrates the period of time. As in Figure \ref{fig10}, the average of write requests in Paxos has an ascending trend. But, Raft still has a regular pattern. A maximum number of write requests in Paxos reaches 14/sec, but it is 11/sec in Raft. This issue again denotes on an unbalanced load of Paxos, rather than Raft. It could be inferred that in Paxos, read and write requests has an inverse relationship with latency, while in Raft requests still follow a less constant pattern. These figures illustrate the write requests in case of using Raft is more organized, and its processing number for the  write requests is up to 11, but the number of write requests in Paxos is heterogeneous in different numbers that can cause increasing pressure on the nodes.

The vertical axis shows the number of read requests in a second in the cluster and the horizontal axis indicates the period of time. According to Figure \ref{fig11}, the number of read requests in Paxos is fluctuating by increasing the system load, while the number of read requests in Raft always is a constant number. These figures show the read requests in case of using Raft is optimized and regarding this fact, read request latency is also inspired. It can be seen from figures of Paxos that the number of read requests somehow are increased.

As the summary of aforementioned issues, we can claim that Raft was always better than Paxos in our measured criteria. The main reason of superiority of the Raft was appropriate load balancing that entailed more flexibility, steadiness, less latency and more speed. All these advantages led to better performance of the whole system.

\section{Conclusion and future work}
\label{sec.con}
The need for scalability and accessibility in distributed database system entailed emergence of consensus algorithms to manage the replicas that provide the scalability and availability for the system; because previous approaches like master-slave and tripartite replications could not solve the problem of the clusters with a large number of nodes. In this paper, we investigated some of the prominent consensus algorithms such as Chandra-Toueg, Paxos, and Raft. Then, by focusing on the effect of read/write request latency on overall performance, we attempted to find the best solution for Cassandra database system. Experiments indicated that Raft had the best performance of other algorithms. So, we tested Raft algorithm on Cassandra database and used Opscenter and Stress tool for monitoring and imposing high load to the system, respectively. Results showed the acceptable load balancing, and consequently improving the efficiency of the system by applying Raft.

Cassandra employs the queues for putting the requests into order. One of the ideas for future work is to inspect and change the strategies of managing the lines to achieve better performance. Also, by partitioning the clusters of Cassandra, improving the efficiency of Raft algorithm can be targeted.

\bibliographystyle{spmpsci}      
\bibliography{ref}   

\begin{thebibliography}{10}
\providecommand{\url}[1]{{#1}}
\providecommand{\urlprefix}{URL }
\expandafter\ifx\csname urlstyle\endcsname\relax
  \providecommand{\doi}[1]{DOI~\discretionary{}{}{}#1}\else
  \providecommand{\doi}{DOI~\discretionary{}{}{}\begingroup
  \urlstyle{rm}\Url}\fi

\bibitem{ref63}
Amid, S., Mesri~Gundoshmian, T.: Prediction of output energy based on different
  energy inputs on broiler production using application of adaptive
  neural-fuzzy inference system.
\newblock Agriculture Science Developments \textbf{5}(2), 14--21 (2016)

\bibitem{ref23}
Anderson, D.P.: Boinc: A system for public-resource computing and storage.
\newblock In: Grid Computing, 2004. Proceedings. Fifth IEEE/ACM International
  Workshop on, pp. 4--10. IEEE (2004)

\bibitem{ref40}
Baker, J., Bond, C., Corbett, J.C., Furman, J., Khorlin, A., Larson, J., Leon,
  J.M., Li, Y., Lloyd, A., Yushprakh, V.: Megastore: Providing scalable, highly
  available storage for interactive services  (2011)

\bibitem{ref31}
Bartoli, A., Calabrese, C., Prica, M., Di~Muro, E.A., Montresor, A.: Adaptive
  message packing for group communication systems.
\newblock In: OTM Confederated International Conferences" On the Move to
  Meaningful Internet Systems", pp. 912--925. Springer (2003)

\bibitem{ref24}
Bertsekas, D.P., Tsitsiklis, J.N.: Parallel and distributed computation:
  numerical methods, vol.~23.
\newblock Prentice hall Englewood Cliffs, NJ (1989)

\bibitem{ref39}
Burrows, M.: The chubby lock service for loosely-coupled distributed systems.
\newblock In: Proceedings of the 7th symposium on Operating systems design and
  implementation, pp. 335--350. USENIX Association (2006)

\bibitem{ref43}
Calder, B., Wang, J., Ogus, A., Nilakantan, N., Skjolsvold, A., McKelvie, S.,
  Xu, Y., Srivastav, S., Wu, J., Simitci, H., et~al.: Windows azure storage: a
  highly available cloud storage service with strong consistency.
\newblock In: Proceedings of the Twenty-Third ACM Symposium on Operating
  Systems Principles, pp. 143--157. ACM (2011)

\bibitem{ref2}
Campbell, D.G., Kakivaya, G., Ellis, N.: Extreme scale with full sql language
  support in microsoft sql azure.
\newblock In: Proceedings of the 2010 ACM SIGMOD International Conference on
  Management of data, pp. 1021--1024. ACM (2010)

\bibitem{ref52}
Cao, M., Morse, A.S., Anderson, B.: Coordination of an asynchronous multi-agent
  system via averaging.
\newblock IFAC Proceedings Volumes \textbf{38}(1), 17--22 (2005)

\bibitem{ref20}
Caron, E., Desprez, F., Tedeschi, C.: Enhancing computational grids with
  peer-to-peer technology for large scale service discovery.
\newblock Journal of Grid Computing \textbf{5}(3), 337--360 (2007)

\bibitem{ref21}
Celaya, J., Arronategui, U.: Distributed scheduler of workflows with deadlines
  in a p2p desktop grid.
\newblock In: 2010 18th Euromicro Conference on Parallel, Distributed and
  Network-based Processing, pp. 69--73. IEEE (2010)

\bibitem{ref3}
Chang, F., Dean, J., Ghemawat, S., Hsieh, W.C., Wallach, D.A., Burrows, M.,
  Chandra, T., Fikes, A., Gruber, R.E.: Bigtable: A distributed storage system
  for structured data.
\newblock ACM Transactions on Computer Systems (TOCS) \textbf{26}(2), 4 (2008)

\bibitem{ref19}
Charr, J.C., Couturier, R., Laiymani, D.: Jacep2p-v2: A fully decentralized and
  fault tolerant environment for executing parallel iterative asynchronous
  applications on volatile distributed architectures.
\newblock Future Generation Computer Systems \textbf{27}(5), 606--613 (2011)

\bibitem{ref18}
Chmaj, G., Walkowiak, K.: A p2p computing system for overlay networks.
\newblock Future Generation Computer Systems \textbf{29}(1), 242--249 (2013)

\bibitem{ref4}
Cooper, B.F., Ramakrishnan, R., Srivastava, U., Silberstein, A., Bohannon, P.,
  Jacobsen, H.A., Puz, N., Weaver, D., Yerneni, R.: Pnuts: Yahoo!'s hosted data
  serving platform.
\newblock Proceedings of the VLDB Endowment \textbf{1}(2), 1277--1288 (2008)

\bibitem{ref41}
Corbett, J.C., Dean, J., Epstein, M., Fikes, A., Frost, C., Furman, J.J.,
  Ghemawat, S., Gubarev, A., Heiser, C., Hochschild, P., et~al.: Spanner:
  Google’s globally distributed database.
\newblock ACM Transactions on Computer Systems (TOCS) \textbf{31}(3), 8 (2013)

\bibitem{ref10}
Dabek, F., Kaashoek, M.F., Karger, D., Morris, R., Stoica, I.: Wide-area
  cooperative storage with cfs.
\newblock ACM SIGOPS Operating Systems Review \textbf{35}(5), 202--215 (2001)

\bibitem{ref5}
DeCandia, G., Hastorun, D., Jampani, M., Kakulapati, G., Lakshman, A., Pilchin,
  A., Sivasubramanian, S., Vosshall, P., Vogels, W.: Dynamo: amazon's highly
  available key-value store.
\newblock ACM SIGOPS Operating Systems Review \textbf{41}(6), 205--220 (2007)

\bibitem{ref33}
Fidge, C.J.: Timestamps in message-passing systems that preserve the partial
  ordering.
\newblock Australian National University. Department of Computer Science (1987)

\bibitem{ref37}
Fischer, M.J., Lynch, N.A., Paterson, M.S.: Impossibility of distributed
  consensus with one faulty process.
\newblock Journal of the ACM (JACM) \textbf{32}(2), 374--382 (1985)

\bibitem{ref28}
Gazi, V.: Stability analysis of swarms.
\newblock Ph.D. thesis, The Ohio State University (2002)

\bibitem{ref62}
Goudarzi, P., Malazi, H.T., Ahmadi, M.: Khorramshahr: A scalable peer to peer
  architecture for port warehouse management system.
\newblock Journal of Network and Computer Applications \textbf{76}, 49--59
  (2016)

\bibitem{ref47}
Guerraoui, R., Levy, R.R., Pochon, B., Qu{\'e}ma, V.: Throughput optimal total
  order broadcast for cluster environments.
\newblock ACM Transactions on Computer Systems (TOCS) \textbf{28}(2), 5 (2010)

\bibitem{ref53}
Hayashibara, N., Urb{\'a}n, P., Schiper, A., Katayama, T.: Performance
  comparison between the paxos and chandra-toueg consensus algorithms.
\newblock In: Proc. Int" l Arab Conference on Information Technology (ACIT
  2002), LSR-CONF-2002-005, pp. 526--533 (2002)

\bibitem{ref6}
Herlihy, M., Rajsbaum, S., Tuttle, M.: An axiomatic approach to computing the
  connectivity of synchronous and asynchronous systems.
\newblock Electronic Notes in Theoretical Computer Science \textbf{230},
  79--102 (2009)

\bibitem{ref11}
Hughes, D., Coulson, G., Walkerdine, J.: Free riding on gnutella revisited: the
  bell tolls?
\newblock IEEE distributed systems online \textbf{6}(6) (2005)

\bibitem{ref42}
Isard, M.: Autopilot: automatic data center management.
\newblock ACM SIGOPS Operating Systems Review \textbf{41}(2), 60--67 (2007)

\bibitem{ref26}
Jadbabaie, A., Lin, J., Morse, A.S.: Coordination of groups of mobile
  autonomous agents using nearest neighbor rules.
\newblock IEEE Transactions on automatic control \textbf{48}(6), 988--1001
  (2003)

\bibitem{ref60}
Junqueira, F.P., Reed, B.C., Serafini, M.: Zab: High-performance broadcast for
  primary-backup systems.
\newblock In: 2011 IEEE/IFIP 41st International Conference on Dependable
  Systems \& Networks (DSN), pp. 245--256. IEEE (2011)

\bibitem{ref49}
Junqueira, F.P., Reed, B.C., Serafini, M.: Zab: High-performance broadcast for
  primary-backup systems.
\newblock In: 2011 IEEE/IFIP 41st International Conference on Dependable
  Systems \& Networks (DSN), pp. 245--256. IEEE (2011)

\bibitem{ref32}
Lamport, L.: Time, clocks, and the ordering of events in a distributed system.
\newblock Communications of the ACM \textbf{21}(7), 558--565 (1978)

\bibitem{ref54}
Lamport, L.: The part-time parliament.
\newblock ACM Transactions on Computer Systems (TOCS) \textbf{16}(2), 133--169
  (1998)

\bibitem{ref36}
Lamport, L., Shostak, R., Pease, M.: The byzantine generals problem.
\newblock ACM Transactions on Programming Languages and Systems (TOPLAS)
  \textbf{4}(3), 382--401 (1982)

\bibitem{ref38}
Lampson, B.: How to build a highly available system using consensus.
\newblock Distributed Algorithms pp. 1--17 (1996)

\bibitem{ref59}
Liskov, B., Cowling, J.: Viewstamped replication revisited  (2012)

\bibitem{ref51}
Liskov, B., Ghemawat, S., Gruber, R., Johnson, P., Shrira, L., Williams, M.:
  Replication in the Harp file system.
\newblock ACM (1991)

\bibitem{ref22}
Litzkow, M.J., Livny, M., Mutka, M.W.: Condor-a hunter of idle workstations.
\newblock In: Distributed Computing Systems, 1988., 8th International
  Conference on, pp. 104--111. IEEE (1988)

\bibitem{ref17}
Lucchese, C., Mastroianni, C., Orlando, S., Talia, D.: Mining@ home: toward a
  public-resource computing framework for distributed data mining.
\newblock Concurrency and Computation: Practice and Experience \textbf{22}(5),
  658--682 (2010)

\bibitem{ref25}
Lynch, N.A.: Distributed algorithms.
\newblock Morgan Kaufmann (1996)

\bibitem{ref48}
Marandi, P.J., Primi, M., Schiper, N., Pedone, F.: Ring paxos: A
  high-throughput atomic broadcast protocol.
\newblock In: 2010 IEEE/IFIP International Conference on Dependable Systems \&
  Networks (DSN), pp. 527--536. IEEE (2010)

\bibitem{ref50}
Moraru, I., Andersen, D.G., Kaminsky, M.: There is more consensus in
  egalitarian parliaments.
\newblock In: Proceedings of the Twenty-Fourth ACM Symposium on Operating
  Systems Principles, pp. 358--372. ACM (2013)

\bibitem{ref27}
Moreau, L.: Leaderless coordination via bidirectional and unidirectional
  time-dependent communication.
\newblock In: Decision and Control, 2003. Proceedings. 42nd IEEE Conference on,
  vol.~3, pp. 3070--3075. IEEE (2003)

\bibitem{ref45}
Nagle, J.: Congestion control in ip/tcp internetworks  (1984)

\bibitem{ref14}
Nezarat, A., Raja, M., Dastghaibifard, G.: A new high performance gpu-based
  approach to prime numbers generation.
\newblock World Applied Programming \textbf{5}(1), 1--7 (2015)

\bibitem{ref8}
Nosrati, M., Nosrati, M., Karimi, R., Karimi, R.: Energy efficient and latency
  optimized media resource allocation.
\newblock International Journal of Web Information Systems \textbf{12}(1),
  2--17 (2016)

\bibitem{ref15}
Okelo, B., Mogotu, P., Omaoro, S., Rwenyo, C.: Generalization in metric space.
\newblock General Scientific Researches \textbf{4}(1), 1--4 (2016)

\bibitem{ref30}
Olfati-Saber, R., Murray, R.M.: Consensus problems in networks of agents with
  switching topology and time-delays.
\newblock IEEE Transactions on automatic control \textbf{49}(9), 1520--1533
  (2004)

\bibitem{ref56}
Ongaro, D.: Consensus: Bridging theory and practice.
\newblock Ph.D. thesis, STANFORD UNIVERSITY (2014)

\bibitem{ref55}
Ongaro, D., Ousterhout, J.: In search of an understandable consensus algorithm.
\newblock In: 2014 USENIX Annual Technical Conference (USENIX ATC 14), pp.
  305--319 (2014)

\bibitem{ref46}
Padmanabhan, V.N., Mogul, J.C.: Improving http latency.
\newblock Computer Networks and ISDN Systems \textbf{28}(1), 25--35 (1995)

\bibitem{ref16}
P{\'e}rez-Miguel, C., Miguel-Alonso, J., Mendiburu, A.: High throughput
  computing over peer-to-peer networks.
\newblock Future Generation Computer Systems \textbf{29}(1), 352--360 (2013)

\bibitem{ref35}
Pregui{\c{c}}a, N., Baquero, C., Almeida, P.S., Fonte, V., Gon{\c{c}}alves, R.:
  Dotted version vectors: Logical clocks for optimistic replication.
\newblock arXiv preprint arXiv:1011.5808  (2010)

\bibitem{ref9}
Sabaghi, M., Dashtbayazi, M., Marjani, S.: Dynamic hysteresis band fixed
  frequency current control.
\newblock World Applied Programming \textbf{6}, 1--4 (2016)

\bibitem{ref1}
Sadalage, P.J., Fowler, M.: NoSQL distilled: a brief guide to the emerging
  world of polyglot persistence.
\newblock Pearson Education (2012)

\bibitem{ref44}
Santos, N., Schiper, A.: Optimizing paxos with batching and pipelining.
\newblock Theoretical Computer Science \textbf{496}, 170--183 (2013)

\bibitem{ref12}
Spitz, D., Hunter, S.D.: Contested codes: The social construction of napster.
\newblock The information society \textbf{21}(3), 169--180 (2005)

\bibitem{ref13}
Stoica, I., Adkins, D., Zhuang, S., Shenker, S., Surana, S.: Internet
  indirection infrastructure.
\newblock ACM SIGCOMM Computer Communication Review \textbf{32}(4), 73--86
  (2002)

\bibitem{ref7}
Tanenbaum, A.S., Van~Steen, M.: Distributed systems.
\newblock Prentice-Hall (2007)

\bibitem{ref29}
Vicsek, T., Czir{\'o}k, A., Ben-Jacob, E., Cohen, I., Shochet, O.: Novel type
  of phase transition in a system of self-driven particles.
\newblock Physical review letters \textbf{75}(6), 1226 (1995)

\bibitem{ref34}
Vinoski, S.: Rediscovering distributed systems.
\newblock IEEE Internet Computing \textbf{18}(2) (2014)

\end{thebibliography}

\end{document}